\documentclass[10pt,
               aps,
               pre,
               twocolumn,
               showpacs, 
               groupedaddress]{revtex4-2}
\usepackage[utf8]{inputenc}
\usepackage{graphicx}
\usepackage{bm}
\graphicspath{{images/}}
\usepackage[caption=false]{subfig}
\usepackage{amsmath, amsfonts, amssymb}
\usepackage[linktocpage=true,
  colorlinks=true, 
  pdfborder={0 0 0},
  linkcolor=blue,
  citecolor=blue,
  filecolor=yellow,
  urlcolor=blue,
  bookmarks,
  pdfauthor={},
]{hyperref}
\usepackage{tikz}
\makeatletter
\newcommand*{\rom}[1]{\expandafter\romannumeral #1}
\makeatother
\begin{document}

\title{Generalized Fokker-Planck equation for the active
Brownian motion}

\author{Sanju S Pillai}
\author{M. Muhsin}



\affiliation{Department of Physics, University of Kerala, Kariavattom, Thiruvananthapuram-$695581$, India}

\author{M. Sahoo}
\email{jolly.iopb@gmail.com}
\affiliation{Department of Physics, University of Kerala, Kariavattom, Thiruvananthapuram-$695581$, India}

\date{\today}

\begin{abstract}
 We investigate the dynamics of an inertial active Ornstein-Uhlenbeck particle suspended in a non-Markovian environment. The particle is additionally subjected to external forces, such as harmonic confinement and a magnetic field.
 Motivated by the importance of understanding the non-Markovian behavior of complex environments, we examine the impact of a viscoelastic medium by employing the Jeffrey fluid framework for modeling the particle motion, which effectively captures both viscous and elastic contributions of the environment.  Within this model, we explicitly derive the corresponding Fokker–Planck equation for each case.
Building on this, we extend the analysis to general non-Markovian framework and derive the corresponding generalized Fokker–Planck equation for a free active particle. Furthermore, we obtain the probability distribution function valid for arbitrary memory kernel under various conditions, including both free and confined motion with and without a magnetic field. To the best of our knowledge, this represents the first attempt to establish such a comprehensive formalism for the probabilistic description of an active particle subjected to non-Markovian memory effects.
This formulation provides a solid basis for analyzing the dynamics of an active particle in a non-Markovian environment, such as mucus and polymer solutions, and further allows the study of relaxation in confined geometries and responses to external fields.

\end{abstract}

\maketitle
\section{INTRODUCTION}\label{sec:intro}

Active matter refers to systems of self-propelled particles that extract energy from their surroundings to generate directed motion, thereby maintaining a state far from equilibrium. Such systems exhibit dynamical behaviors beyond the scope of equilibrium statistical mechanics ~\cite{Ramaswamy2017,Bechinger2016,Gompper2020,Pietzonka2021,DeMagistris2015}. Examples of active matter range from bacterial colonies~\cite{aranson2022bacterial} and molecular motors~\cite{reimann2002brownian} to synthetic Janus particles and active colloids \cite{Walther2013,Howse2007}. Unlike passive systems, active particles display persistent motion and long-lived correlations, often giving rise to emergent phenomena such as clustering, swarming, and motility-induced phase separation~\cite{gompper2025active}. 
Several theoretical models have been proposed to capture the essential features of active particles.
One widely used framework is that of active Brownian particles (ABPs), which describes propulsion with rotational diffusion and is widely used in synthetic and biological contexts of active matter~\cite{Howse2007, Fily2019, Cates2011}.
Another important class is the Run-and-Tumble particles (RTPs), which are often used to model the intermittent propulsion characteristic of bacterial motility \cite{berg1972chemotaxis, martens2012probability}.
In addition to these, the active Ornstein–Uhlenbeck particles (AOUPs) describe self-propulsion through a stochastic force with an exponential temporal correlation
\cite{Lehle2018,Bonilla2019,Martin2021,Caprini2019,Caprini2021,Dabelow2019,Berthier2019,Wittmann2018,Fily2019,mandal2017entropy,fodor2016nonequilibrium,muhsin2021orbital}. However, most of these frameworks assume a Markovian environment, where the bath in which the particle is suspended is memoryless, and the dynamics depend only on the instantaneous state of the particle.

In reality, however, many active systems operate in non-Markovian environments, where the motion of the particle reflects past interactions with the medium \cite{Marchetti2013,Cates2011}. Such environments are common in biological and soft-matter systems, which often possess structural complexity and finite relaxation times \cite{Winkler2020}. 
For example, bacteria that move through mucus~\cite{mirbagheri2016helicobacter}, colloids suspended in polymeric solutions~\cite{plan2020active}, or molecular motors that traverse the crowded cytoplasm experience memory effects. 
These effects manifest themselves as viscoelastic responses, where the medium dissipates and stores energy, leading to long-term correlations, anomalous diffusion, and nontrivial relaxation phenomena \cite{Zhang2019,Vainstein2005,Vitali2022, Metzler2000, DelCastilloNegrete2004, adersh2025transition}. Importantly, while Markovian models capture many aspects of active dynamics, they fail to reproduce these nonequilibrium signatures that are inherently tied to non-Markovian environments~\cite{luczka2005nonmarkov, wisniewski2024memory}. Understanding how activity couples with viscoelasticity is therefore essential for developing accurate theoretical frameworks that bridge microscopic stochastic dynamics with macroscopic observables.
Non-Markovian dynamics can be described using the generalized Langevin equation, which gives a trajectory-based account of how the position of the particle evolves randomly over time~\cite{igor2012visco, sevilla2019generalized}.
In contrast, the generalized Fokker–Planck equation provides a probabilistic framework~\cite{adelman1976fokker, Das2017fokker}.
It describes how the probability distribution of the particle changes with time.
Such a framework would allow one to describe the macroscopic behavior of the system much more effectively and also allow for the calculation of moments.
However, a probabilistic framework for an active particle in non-Markovian environments has not been extensively explored.
In this work, we focus on developing such a probabilistic description of the system by deriving the generalized Fokker-Planck equation (FPE) and computing the distribution function of an active particle in a non-Markovian environment.

Moreover, understanding the dynamics of confined active matter is crucial for a wide range of applications, including control of microswimmers in optical traps \cite{Zhang2021,Kamath2023}, regulation of intracellular transport processes under mechanical constraints \cite{Buttinoni2022,Takatori2016}, and design of artificial colloids operating in structured or engineered environments~\cite{tkachenko2023evanescent}. Similarly, introducing an external magnetic field in such a system can provide an additional degree of control, enabling the directed motion of magnetically active particles~\cite{nourhani2021Spontaneous, cifmmode2006dynamics, tierno2008magnetically}. Such mechanisms have significant potential for biomedical applications, such as targeted drug delivery~\cite{medina2016cellular, liu2025supramolecular, pankhurst2003applications}, microfluidic transport~\cite{zou2018composite}, cell sorting, and bioseparation techniques~\cite{ayse2009label}.
Motivated by this, we have investigated the non-Markovian dynamics of an active particle in various settings, including those with and without confinement, as well as in the presence or absence of an external magnetic field. Using a simplified model for a viscoelastic environment (i.e., the Jeffrey fluid framework), we derive the Fokker-Planck equation associated with the dynamics of an active particle. Further, we have derived the generalized Fokker-Planck equation and the corresponding probability distribution function for a free particle suspended in a non-Markovian environment. By considering an arbitrary friction kernel, we employ an analytic formalism to obtain the probability distribution for all the cases considered.
Our study provides a versatile framework that extends the applicability of active matter to more realistic environments. This work thus contributes to the broader understanding of active matter in complex media, with implications for the physics of soft matter, biological transport, and the design of active materials.

\section{model}
We consider the motion of an inertial active Ornstein-Uhlenbeck particle suspended in a viscoelastic bath and confined by a two-dimensional potential $V(x,y)$. The particle is additionally subjected to an external magnetic field $\mathbf{B} = B \hat{k}$. 
The viscoelastic properties of the medium induce memory effects, leading to non-Markovian dynamics of the particle. Hence, the dynamics of the particle can be described by the generalized Langevin equation of motion.
\begin{equation}
m\mathbf{\ddot  r} = -\int_0^t \gamma(t-t')\mathbf{\dot r}(t') dt' - \nabla V + q (\mathbf{\dot r} \times \mathbf{B}) + \boldsymbol{\eta}(t) + \boldsymbol{\Omega}(t).
\label{GLE}
\end{equation}

Here, the position vector of the particle in the \(x\)-\(y\) plane is given by \( \mathbf{r}(t) = x(t) \hat{i} + y(t) \hat{j} \). The \( m \) denotes the mass of the particle. 
The first term on the right-hand side of the above equation represents the viscoelastic drag force, the second term accounts for the conservative force from the external potential, and the third term describes the non-conservative force due to the applied magnetic field.
The fourth term $\boldsymbol{\eta}(t)$ represents thermal fluctuations, while the last term $\boldsymbol{\Omega}(t)$ corresponds to the active force modeled by an Ornstein-Uhlenbeck process. The noise $\boldsymbol{\eta}(t)$ is governed by specific statistical properties $\langle \eta_i(t) \rangle = 0$ and
\begin{equation}
\quad \langle \eta_i(t) \eta_j(t') \rangle = \delta_{ij} k_B T \gamma (t - t'), 
\label{eta_correlation}
\end{equation}
for \( i,j \in \{x,y\} \).
Here, $\gamma(t-t')$ represents the friction kernel. In this work, we consider Jeffreys' fluid model~\cite{}, for which $\gamma(t-t')$ is given by 
\begin{equation}
\gamma(t-t') = \frac{\gamma_f}{2} \delta(t-t') + \frac{\gamma_s}{2 t_{s}} e^{-(t-t')/2 t_{s}}.
\label{Jeff_fluid}
\end{equation}
This model captures both viscous and elastic contributions, offering a conceptually intuitive and analytically tractable framework for analysis. The first term of Eq.~\eqref{Jeff_fluid} represents the viscous response of the medium, which follows a delta-correlated kernel, while the second term of Eq.~\eqref{Jeff_fluid} is the elastic component that exhibits a mono-exponential decay. 
The \( \gamma_{f} \) is the viscous coefficient that quantifies the viscous effects of the bath on the particle, while the parameters \( \gamma_s \) and \( t_{s} \) govern the elastic response. A higher value of \( t_{s} \) indicates slower relaxation of fluid particles, while \( \gamma_s \) determines the strength of the relaxation dynamics that influences the motion of the particles. In the viscous limit, either by setting \( \gamma_s = 0 \) or letting \( t_{s} \to 0 \), the memory effects are eliminated, reducing the system to purely Markovian dynamics. For \( \gamma_s = 0 \), the elastic contribution vanishes, leaving only the delta-correlated viscous term 
\begin{equation}
\lim_{\gamma_s\to 0}\gamma(t - t') = \frac{\gamma_{f}}{2} \delta(t - t').
\end{equation}
This equation recovers the classical Langevin dynamics described in a purely viscous medium. Alternatively, in the limit \( t_{s} \to 0 \), the exponential decay term in the memory kernel transforms into a delta function  
\begin{equation}
\lim_{t_c\to 0} \gamma(t-t') =  \frac{\gamma_f + \gamma_s}{2} \delta(t - t').
\end{equation}
In both these discussed limits, the system behaves as a purely viscous medium, recovering the classical Langevin framework.

The term \(\boldsymbol{\Omega}(t)\) in Eq.~\eqref{GLE} accounts for non-thermal fluctuations, following an Ornstein-Uhlenbeck (OU) process described by
\begin{equation}
t_{c} \boldsymbol{\dot{\Omega}} = -\boldsymbol{\Omega} + D_\Omega  \boldsymbol{\zeta}(t).
\label{active_dynamics}
\end{equation}

where \( \boldsymbol{\Omega}(t) \) is a Gaussian process with zero mean and exhibits exponential temporal correlations given by,
\begin{equation}
\langle \Omega_i(t) \rangle = 0, \quad \langle \Omega_i(t) \Omega_j(t') \rangle = \delta_{ij} D_\Omega^2 e^{-(t-t')/t_{c}},
\end{equation}
for \( i,j \in \{x,y\} \).
The correlation decays exponentially with a characteristic time scale \( t_{c} \), representing the persistence of self-propulsion or the system activity time scale. The parameter \( D_\Omega \) denotes the magnitude of the active driving force, while \( \boldsymbol{\zeta}(t) \) is a white noise process with delta-correlated fluctuations.  

\section{result and discussion}

\subsection{FREE ACTIVE PARTICLE}
In this section, we consider a free inertial active Ornstein-Uhlenbeck particle in a viscoelastic bath. 
Since the particle is unbounded, the potential is zero, and in the absence of an external magnetic field, no non-conservative force acts on it. 
Consequently, the Generalized Langevin Equation, Eq.~\eqref{GLE} takes the form
\begin{equation}
	m \boldsymbol{\dot{v}} = - \int_{0}^{t} \gamma(t - t') \boldsymbol{v}(t') dt' + \boldsymbol{\eta}(t) + \boldsymbol{\Omega}(t).
	\label{GLE_1}
\end{equation}
In Eq.~\eqref{GLE_1}, $\boldsymbol{\eta}(t)$ represents a Gaussian thermal noise. 
This noise term is intrinsically related to the time-dependent damping strength $\gamma(t-t')$ through the fluctuation-dissipation theorem as in Eq.~\eqref{eta_correlation}. 
First, we model the medium as a Jeffreys fluid as in Eq.~\eqref{Jeff_fluid}.
We decompose the noise term $\boldsymbol{\eta}(t)$  into two distinct contributions: one representing the viscous component and the other accounting for the exponentially decaying elastic part as
\begin{equation}
	\boldsymbol{\eta}(t) = \boldsymbol{\eta_{1}}(t) + \boldsymbol{\eta_{2}}(t).
\end{equation}
Accordingly, the correlations of these two independent noise terms are given by
\begin{equation}
	\langle \eta_{1i}(t) \cdot \eta_{1j}(t') \rangle =  \frac{k_B T\gamma_f}{2} \delta_{ij} \delta(t - t'),
\end{equation}
and
\begin{equation}
	\langle \eta_{2i}(t) \cdot \eta_{2j}(t') \rangle =  \frac{k_B T\gamma_s}{2 t_{s}} \delta_{ij} e^{-\frac{(t - t')}{t_{s}}}.
\end{equation}
Here, $\boldsymbol{\eta_2}$ corresponds to the noise contribution due to the elastic kernel part, and it follows the dynamics given by

\begin{equation}
	\boldsymbol{\dot{\eta_2}} = -\frac{1}{t_s} \boldsymbol{\eta_2} + \sqrt{\frac{k_B T  \gamma_s}{t_s}} \boldsymbol{\zeta_1},
\end{equation}
where,
\begin{equation}
	\langle \zeta_{1i}(t) \zeta_{1j}(t') \rangle =  \delta_{ij} \delta(t - t').
\end{equation}
By incorporating the above mentioned decomposed noise terms and the friction kernel given by Eq.~\eqref{Jeff_fluid} into Eq.~\eqref{GLE_1}, we obtain

\begin{equation}
	m \boldsymbol{\dot{v}} = -\frac{\gamma_f}{2} \boldsymbol{v} - \int_{0}^{t} \frac{\gamma_s}{2 t_s} e^{-\frac{(t - t')}{t_s}} \boldsymbol{v}(t') \, dt' + \boldsymbol{\eta_1}(t) + \boldsymbol{\eta_2}(t) + \boldsymbol{\Omega}(t).
\end{equation}
We can proceed by substituting the elastic integral term as
\begin{equation}
	\boldsymbol{\xi}(t) = \frac{1}{t_s} \int_{0}^{t}  e^{-\frac{(t - t')}{t_s}} \boldsymbol{v}(t') \, dt'.
\end{equation}
On taking the derivative of the above equation following Leibniz's rule, we obtain the dynamics of the variable $\boldsymbol{\xi}$ as
\begin{equation}
	\boldsymbol{\dot{\xi}} = -\frac{1}{t_s} \boldsymbol{\xi} + \frac{1}{t_s} \boldsymbol{v}.
\end{equation}
The active force $\boldsymbol{\Omega}(t)$ follows the dynamics given by Eq.~\eqref{active_dynamics} with a similar noise term $\boldsymbol{\zeta_2}$. Now, the Eq.~\eqref{GLE_1} can be expressed as a system of equations given by
	\begin{align}
		\boldsymbol{\dot{v}} &= -\frac{\gamma_f}{2m} \boldsymbol{v} - \frac{\gamma_s}{2m} \boldsymbol{\xi} + \frac{1}{m} \boldsymbol{\eta_1} + \frac{1}{m} \boldsymbol{\eta_2} + \frac{1}{m} \boldsymbol{\Omega} \\
		\boldsymbol{\dot{\xi}} &= \frac{1}{t_s} \boldsymbol{v} - \frac{1}{t_s} \boldsymbol{\xi} \\
		\boldsymbol{\dot{\eta_2}} &= -\frac{1}{t_s} \boldsymbol{\eta_2} + \sqrt{\frac{k_B T \gamma_s}{t_s}} \boldsymbol{\zeta_1}\\
		\boldsymbol{\dot{\Omega}} &= -\frac{1}{t_c} \boldsymbol{\Omega} + \sqrt{\frac{2 D_{\Omega}^{2}}{t_c}} \boldsymbol{\zeta_2}.
	\end{align}
	\begin{widetext}
		\noindent
		The above set of equations can be represented in a matrix form as
		\begin{equation}
			\dot{\boldsymbol{X}} = A \boldsymbol{X} + G \boldsymbol{\eta}(t),
			\label{eq:dyn_matrix_form}
		\end{equation}
		where \( A \) is the drift matrix that governs the deterministic dynamics of the system, and \( G \) is the noise-coupling matrix that determines how the stochastic noise \( \boldsymbol{\eta}(t) \) influences each component of the system. The matrix product \( B = G G^T \) defines the diffusion matrix, which characterizes the strength and anisotropy of the stochastic fluctuations in the system. The expressions for these matrices are given by 
		\begin{equation}
			\boldsymbol{X} =
			\begin{bmatrix}
				\boldsymbol{v} \\ \boldsymbol{\xi} \\ \boldsymbol{\eta_2} \\ \boldsymbol{\Omega}
			\end{bmatrix}
			, \quad
			A =
			\begin{bmatrix}
				
				0 & -\frac{\gamma_f}{2m} & -\frac{\gamma_s}{2m} & \frac{1}{m} & \frac{1}{m} \\
				0 & \frac{1}{t_{s}} & -\frac{1}{t_{s}} & 0 & 0 \\
				0 & 0 & 0 & -\frac{1}{t_{s}} & 0 \\
				0 & 0 & 0 & 0 & -\frac{1}{t_{c}}
			\end{bmatrix}, \quad
			B =
			\begin{bmatrix}
				
				0 & \frac{1}{m^2} & 0 & 0 & 0 \\
				0 & 0 & 0 & 0 & 0 \\
				0 & 0 & 0 & \frac{k_B T \gamma_s}{t_{s}} & 0 \\
				0 & 0 & 0 & 0 & \frac{2 D_{\Omega}^{2}}{t_{c}}
			\end{bmatrix}, \  \text{and}\quad
			\boldsymbol{\eta} = 
			\begin{bmatrix}
				\boldsymbol{\eta_1} \\ 0 \\ \boldsymbol{\zeta_1} \\ \boldsymbol{\zeta_2}
			\end{bmatrix}.
		\end{equation}
		With these matrices,
		the general form of the Fokker-Planck equation, which describes the time evolution of the probability density function \(P(\boldsymbol{X}; t)\), is given by~\cite{vankampen2007stochastic},
		\begin{equation}
			\frac{\partial P}{\partial t} = - \nabla \cdot (A \boldsymbol{X} P) + \frac{1}{2} \nabla \cdot (B \nabla P).
			\label{eq:FPE_matrix_form}
		\end{equation}
		Thus, the FPE takes the form,
		\begin{equation}
			\begin{aligned}
				\frac{\partial P}{\partial t} &=  
				\frac{\gamma_f}{2m} \frac{\partial}{\partial \boldsymbol{v}} (\boldsymbol{v} P) 
				+ \left(\frac{\gamma_s \boldsymbol{\xi}}{2m} - \frac{\boldsymbol{\eta_2}}{m} - \frac{\boldsymbol{\Omega}}{m} \right) \frac{\partial P}{\partial \boldsymbol{v}}  - \frac{\boldsymbol{v}}{t_{s}} \frac{\partial P}{\partial \boldsymbol{\xi}} 
				+ \frac{1}{t_{s}} \frac{\partial}{\partial \boldsymbol{\xi}} (\boldsymbol{\xi} P) 
				+ \frac{1}{t_{s}} \frac{\partial}{\partial \boldsymbol{\eta_2}} (\boldsymbol{\eta_2} P) \\
				&\quad + \frac{1}{t_{c}} \frac{\partial}{\partial \boldsymbol{\Omega}} (\boldsymbol{\Omega} P) 
				+ \frac{1}{2m^2} \frac{\partial^2 P}{\partial \boldsymbol{v}^2}  + \frac{k_B T \gamma_s}{2 t_{s}} \frac{\partial^2 P}{\partial \boldsymbol{\eta_2}^2} 
				+ \frac{\xi_0^2}{t_{c}} \frac{\partial^2 P}{\partial \boldsymbol{\Omega}^2}.
				\label{jeffry_active_fpe_case1}
			\end{aligned}
		\end{equation}
		
	\end{widetext}
	
	Next, we consider a general non-Markovian memory kernel $\gamma(t-t')$. In this case, the equation of motion [Eq.~\eqref{GLE}] takes the form
	\begin{align}
			\boldsymbol{\dot v}(t)&=-\frac{1}{m}\int_{0}^{t} \gamma(t-t')\boldsymbol{v}(t') dt'+\frac{1}{m}\boldsymbol{\eta}(t)+\frac{1}{m}\boldsymbol{\Omega}(t),
			\label{eq:vdot_general}
		\end{align}
		with
		\begin{align}
			\boldsymbol{\dot \Omega }(t) &= - \frac{1}{t_c} \boldsymbol{\Omega}(t) + D_\Omega \sqrt{\frac{2}{t_c}} \boldsymbol{\zeta}(t).
			\label{eq:Odot_general}
	\end{align}
	
	Taking Laplce tranform of Eqs.~\eqref{eq:vdot_general} and ~\eqref{eq:Odot_general}, and rearranging the terms, we get
	\begin{align}
		&\begin{aligned}
			\boldsymbol{\tilde v}(s) &= \tilde \chi_1(s)\boldsymbol{ v_0}+\tilde \chi_1(s)\tilde \chi_2(s) \boldsymbol{\Omega_0} + \frac{1}{m}\tilde \chi_1(s)  \boldsymbol{\tilde \eta}(s)+\\
			&+\frac{D_\Omega}{m}\sqrt{\frac{2}{t_c}} \tilde \chi_1(s)  \tilde \chi_2(s) \boldsymbol{\tilde\zeta}(s),
			\label{eq:vs_general}
		\end{aligned}
		\\
		&\boldsymbol{
			\tilde \Omega}(s) = \tilde \chi_2(s) \boldsymbol{\Omega_0} + D_\Omega \sqrt{\frac{2}{t_c}} \tilde \chi_2(s)\boldsymbol{\tilde\zeta}(s),
		\label{eq:Os_general}
	\end{align}
	with
	\begin{equation}
		\begin{aligned}
			\tilde\chi_1(s)=&\frac{1}{s+\frac{1}{m}\tilde\gamma(s)} \quad \text{and} \quad \tilde\chi_2(s)=\frac{1}{1+\frac{1}{t_c}},
			\label{response_case1}    
		\end{aligned}
	\end{equation}
	where
	\begin{align}
		\boldsymbol{\Tilde{v}}(s) &= \int\limits_{0}^{\infty} \boldsymbol{v}(t) e^{-st}\; dt,\\
		\boldsymbol{\Tilde{\Omega}}(s) &= \int\limits_{0}^{\infty} \boldsymbol{\Omega}(t) e^{-st}\; dt.
	\end{align}
	The fluctuations in the dynamics can be characterized by introducing two functions $\boldsymbol{g_1}(t)$ and $\boldsymbol{g_2}(t)$ such that
	\begin{align}
		\boldsymbol{g_1}(t) &= \boldsymbol{v}(t)-\chi_1(t)\boldsymbol{v_0}-\chi_{12}(t)\boldsymbol{\Omega_0}, \\
		\boldsymbol{g_2}(t) &= \boldsymbol{\Omega}(t)-\chi_2(t)\boldsymbol{\Omega_0},
	\end{align}
	with $\chi_{12}(t) = (\chi_1 * \chi_2)(t)$, where `$*$' represent the convolution.
	Now taking the inverse Laplace transform of Eqs.~\eqref{eq:vs_general} and ~\eqref{eq:Os_general}, we get the fluctuation terms $\boldsymbol{g_1}(t)$ and $\boldsymbol{g_2}(t)$ as
	\begin{equation}
		\begin{aligned}
			\boldsymbol{g_1}(t) &= \frac{1}{m} \int_0^t \chi_1(t')\boldsymbol{\eta}(t-t') dt'\\&+\frac{D_\Omega}{m}\sqrt{\frac{2}{t_c}}\int_0^t \chi_{12}(t')\boldsymbol{\zeta}(t-t') dt'.\\
			\boldsymbol{g_2}(t) &= {D_\Omega}\sqrt{\frac{2}{t_c}}\int_0^t \chi_2(t')\boldsymbol{\zeta}(t-t') dt'. 
			\label{fluctuation_Case1}
		\end{aligned}
	\end{equation}
	From Eq.~\eqref{response_case1}, we have 
	\begin{equation}
		\begin{aligned}
			\tilde\chi_1(s)\left[s+\frac{1}{m}\tilde\gamma(s)\right]=1,\\
			s\tilde\chi_1(s)-1+\frac{1}{m}\tilde\gamma(s)=0.
			\label{case1:chi_1}
		\end{aligned}    
	\end{equation}
	Also, from Eq.~\eqref{fluctuation_Case1}, it is clear that when $t=0$,  $\boldsymbol{g_1} = 0$ and $\boldsymbol{v}= \boldsymbol{v_0}$, which implies $\chi_1(t=0)=1$. Thus Eq.~\eqref{case1:chi_1} becomes
	\begin{equation}
		\dot\chi_1(t)+\int_0^t \gamma(t-t')\chi_1(t')dt'=0.
	\end{equation}
	Similarly, from Eq.~\eqref{response_case1}, we have
	\begin{equation}
		\begin{aligned}
			\chi_2(t)&=e^{-\frac{t}{t_c}}, \quad \text{and}\\
			\dot\chi_2(t)&=-\frac{1}{t_c} e^{-\frac{t}{t_c}}.
		\end{aligned}
	\end{equation}
	Now, in Eq.~\eqref{fluctuation_Case1}, consider $\chi_{12}(t)$ as
	\begin{equation}
		\chi_{12}(t)=(\chi_1*\chi_2)(t)=\int_0^t \chi_1(t') e^{-\frac{t-t'}{t_c}}dt'.
		\label{eq:chi12_case1}
	\end{equation}
	Taking derivative of $\chi_{12}(t)$ in Eq.~\eqref{eq:chi12_case1}, we get
	\begin{equation}
		\dot\chi_{12}=\chi_1(t)-\frac{1}{t_c} \chi_{12}(t).
	\end{equation}
	
	We now examine the second-order statistical moments related to the components of fluctuations. These moments can be written as a matrix 
	$\Xi(t)$, whose elements are given by $\Xi_{lm}=\langle\boldsymbol{g_l}(t)\cdot\boldsymbol{g_m}(t)\rangle$, where
	$l,m \in \{v,\Omega\}$.
	The matrix $A(t)$ can be written as
	\begin{equation}
		\Xi(t) = 
		\begin{bmatrix}
			\Xi_{11}(t) & \Xi_{12}(t)\\
			\Xi_{21}(t) & \Xi_{22}(t)
		\end{bmatrix},
	\end{equation}
	with
	\begin{equation}
		\Xi_{11}(t) = \frac{K_BT}{m^2} C_\eta(t)+\frac{2D_\Omega^2}{m^2 t_c} \int_0^t \chi_{12}^2(t-t')dt',
	\end{equation}
	where
	\begin{equation}
		C_\eta(t) = \int_0^t dt'\int_0^tdt''\chi_{1}(t-t')\chi_{1}(t-t'')\gamma(|t'-t''|).
	\end{equation}
	Similarly, other components of $A(t)$ are given by
	\begin{equation}
		\Xi_{22}(t)=\frac{2D_\Omega^2}{t_c} \int_0^t\chi_2^2(t-t')dt',
	\end{equation}
	and
	\begin{equation}
		\Xi_{12}(t)=\Xi_{21}(t)=\frac{2D_\Omega^2}{m t_c} \int_0^t \chi_{12}(t-t')\chi_2(t-t')dt'.
	\end{equation}
	Since the equation of motion [Eq.~\eqref{eq:vdot_general}] is linear with the Gaussian noise, the phase space distribution function ($P(\boldsymbol{v},\boldsymbol{\Omega};t)$) can be
	written using the matrix $\Xi(t)$ and its inverse $\Xi^{-1}(t)$ as
	
	\begin{equation}
		P(\boldsymbol{v},\boldsymbol{\Omega};t)=\left(\frac{1}{2 \pi}\right) \frac{1}{\sqrt{|\Xi(t)|}} \exp\left[-\frac{1}{2} \boldsymbol{g}^T(t) \Xi^{-1}(t)\boldsymbol{g}(t)\right],
		\label{Probability_case1}
	\end{equation}
	with
	\begin{equation}
		\boldsymbol{g}(t)=\begin{bmatrix}
			\boldsymbol{g_1}(t)\\
			\boldsymbol{g_2}(2)
		\end{bmatrix}.
	\end{equation}
	Following the expression of FPE from Eq.~\eqref{jeffry_active_fpe_case1}, one can propose the FPE for the non-Markovian dynamics as
	\begin{equation}
		\begin{aligned}
			\frac{\partial P}{\partial t} &= Q_1(t) \frac{\partial}{\partial \boldsymbol{v}}(\boldsymbol{v}P)+Q_2(t) \frac{\partial}{\partial \boldsymbol{v}}(\Omega P)+Q_3(t) \frac{\partial ^2P}{\partial \boldsymbol{v}^2}\\&+Q_4(t) \frac{\partial}{\partial \boldsymbol{\Omega}}(\boldsymbol{\Omega }P)+Q_5(t) \frac{\partial^2P}{\partial \boldsymbol{\Omega}^2}.
		\end{aligned}
		\label{eq:FPE_general_proposed}
	\end{equation}
	
	\begin{widetext}
		The quantities $Q_1(t)$, $Q_2(t)$, $Q_3(t)$, $Q_4(t)$, and $Q_5(t)$   represent time-dependent coefficients that accurately capture the non-Markovian nature of the system. The coefficients can be obtained by substituting the expression of $P(\boldsymbol{v},\boldsymbol{\Omega};t)$ from Eq.~\eqref{Probability_case1} in Eq.~\eqref{eq:FPE_general_proposed} and comparing each terms of both sides, we obtain
		\begin{equation}
			Q_1(t) = 
			\frac{ m \Omega_0 \, \chi_{2}(t) \, \dot{\Xi}_{12}(t) 
				- \Xi_{22}(t)\dot{H}(t) 
				- m \Omega_0 \Xi_{12}(t) \, \dot{\chi}_{2}(t)}
			{G(t)} ,
			\label{eq:Q1_expression}
		\end{equation}
		\begin{equation}
			Q_2(t) = 
			\frac{-\Big( H(t) \dot{\Xi}_{12}(t) 
				- \Xi_{12}(t) H(t) \frac{\dot{\chi}_{2}(t)}{\chi_{2}(t)} \Big)}
			{G(t)} ,
		\end{equation}
		\begin{equation}
			\begin{split}
				Q_3(t) = & \frac{\left[\Xi_{22}(t)H(t)-m\Omega_0 \Xi_{12}(t)\chi_2(t)\right]\dot{\Xi}_{11}(t)}{2G(t)} - \frac{\left[ \Xi_{12}(t)H(t) - m\Omega_0\Xi_{11}(t)\chi_2(t) \right] \dot{\Xi}_{12}(t)}{G(t)} \\
				& + \frac{\left[ \Xi_{12}(t)2 - \Xi_{11}(t)\Xi_{22}(t) \right] \dot{H}(t)}{G(t)} - \frac{\Xi_{12}(t)\left[ m \Omega_0 \Xi_{11}(t)\chi_2(t) - \Xi_{12}(t) H(t) \right] \dot{\chi}_2(t)}{G(t) \chi_2(t)},
			\end{split}
		\end{equation}
		\begin{equation}
			Q_4(t) = -\frac{\dot{\chi}_{2}(t)}{\chi_{2}(t)},
		\end{equation}
		\begin{equation}
			Q_5(t) = \frac{\dot{\Xi}_{22}(t)}{2} - \frac{\Xi_{22}(t) \, \dot{\chi}_{2}(t)}{\chi_{2}(t)},
			\label{eq:Q5_expression}
		\end{equation}
	\end{widetext}
	with
	\begin{equation}
		H(t) = m v_0 \, \chi_{1}(t) + \Omega_0 \, \chi_{12}(t),
	\end{equation}
	and 
	\begin{equation}
		G(t) = \Xi_{22}(t) H(t) - m \Omega_0 \Xi_{12}(t) \, \chi_{2}(t).
	\end{equation}
	Thus, Eq.~\eqref{eq:FPE_general_proposed} represents the FPE for a free active particle in a non-Markovian environment with an arbitrary memory kernel. 
	It can be noted that in the Markovian limit, that is, when $\gamma(t - t') = 2\gamma\delta(t-t')$, the Eqs.~\eqref{eq:Q1_expression}-~\eqref{eq:Q5_expression} reduces to
	\begin{equation}
		\begin{split}
			Q1 &= \frac{\gamma}{m}, \quad Q2 = -\frac{1}{m}, \quad Q3 = \frac{k_B T \gamma}{m^2}, \\
			Q4 &= \frac{1}{t_c}, \quad \text{and} \quad Q5 = \frac{D_\Omega^2}{t_c}.
		\end{split}
		\label{eq:Qi_markov}
	\end{equation}
	Substituting $Q_1$, $Q_2$, $Q_3$, and $Q_4$ from Eq.~\eqref{eq:Qi_markov} in  Eq.~\eqref{eq:FPE_general_proposed}, one can obtain the FPE for an active Brownian particle in the Markovian limit~\cite{muhsin2023ratchet}.
	In the next section, we discuss the case of an active particle in the presence of a harmonic potential.
\subsection{ACTIVE PARTICLE WITH HARMONIC CONFINEMENT}
Here, we consider the case of an active Ornstein-Uhlenbeck particle confined by a harmonic potential with harmonic frequency $\omega$ and suspended in a non-Markovian medium. As discussed in the previous case, first we consider the viscoelastic medium by modeling it as a Jeffrey fluid. The dynamics of the particle [Eq.~\eqref{GLE}] is given by
\begin{equation}
	\begin{aligned}
		m \ddot{\boldsymbol{r}} &= -\frac{\gamma_f}{2} \boldsymbol{\dot{r}} - \int_{0}^{t} \frac{\gamma_s}{2 t_s} e^{-(t-t')/t_s} \boldsymbol{\dot{r}}(t') \, dt' \\
		&\quad + \boldsymbol{\eta(t)} + \boldsymbol{\Omega(t)} - m \omega^2 \boldsymbol{r}.
		\label{eq:case_2_jeffry}
	\end{aligned}
\end{equation}
In order to represent Eq.~\eqref{eq:case_2_jeffry} in matrix form as in Eq.~\eqref{eq:dyn_matrix_form}, the matrices $A$, $B$, and $\bm{X}$ take the form as
\begin{widetext}
	\begin{equation}
		\bm{X} =
		\begin{bmatrix}
			\boldsymbol{r} \\ \boldsymbol{v} \\ \boldsymbol{\xi} \\ \boldsymbol{\eta_2} \\ \boldsymbol{\Omega}
		\end{bmatrix}, \quad
		A =
		\begin{bmatrix}
			0 & 1 & 0 & 0 & 0 \\
			-\omega^2 & -\frac{\gamma_f}{2m} & -\frac{\gamma_s}{2m} & \frac{1}{m} & \frac{1}{m} \\
			0 & \frac{1}{t_{s}} & -\frac{1}{t_{s}} & 0 & 0 \\
			0 & 0 & 0 & -\frac{1}{t_{s}} & 0 \\
			0 & 0 & 0 & 0 & -\frac{1}{t_{c}}
		\end{bmatrix}, \quad 
		\text{and} \quad
		B =
		\begin{bmatrix}
			0 & 0 & 0 & 0 & 0 \\
			0 & \frac{1}{m^2} & 0 & 0 & 0 \\
			0 & 0 & 0 & 0 & 0 \\
			0 & 0 & 0 & \frac{K_B T \gamma_s}{t_{s}} & 0 \\
			0 & 0 & 0 & 0 & \frac{2 \xi_0^2}{t_{c}}.
		\end{bmatrix}.
	\end{equation}
	Consequently, the FPE associated with Eq.~\eqref{eq:case_2_jeffry} takes the form
	\begin{equation}
		\begin{aligned}
			\frac{\partial P}{\partial t} &= 
			- v \frac{\partial P}{\partial \boldsymbol{r}} 
			+ \frac{\gamma_f}{2m} \frac{\partial}{\partial \boldsymbol{v}} (\boldsymbol{v} P) + \left(\omega^2 x +\frac{\gamma_s}{2m} \boldsymbol{\xi} - \frac{\boldsymbol{\eta_2}}{m} - \frac{\boldsymbol{\Omega}}{m} \right) \frac{\partial P}{\partial \boldsymbol{v}}  - \frac{\boldsymbol{v}}{t_{s}} \frac{\partial P}{\partial \boldsymbol{\xi}} 
			+ \frac{1}{t_{s}} \frac{\partial}{\partial \boldsymbol{\xi}} (\boldsymbol{\xi} P) 
			+ \frac{1}{t_{s}} \frac{\partial}{\partial \boldsymbol{\eta_2}} (\boldsymbol{\eta_2} P) \\
			&\quad + \frac{1}{t_{c}} \frac{\partial}{\partial \boldsymbol{\Omega}} (\boldsymbol{\Omega} P) 
			+ \frac{1}{2m^2} \frac{\partial^2 P}{\partial \boldsymbol{v}^2} + \frac{K_B T \gamma_s}{2 t_{s}} \frac{\partial^2 P}{\partial \boldsymbol{\eta_2}^2}
			+ \frac{\xi_0^2}{t_{c}} \frac{\partial^2 P}{\partial \boldsymbol{\Omega}^2}.
		\end{aligned}
		\label{jeffry_active_fpe_case2}
	\end{equation}    
\end{widetext}

Now, we consider the general non-Markovian kernel $\gamma(t - t')$, characterizing a general non-Markovian medium. In this case, the equation of motion [Eq.~\eqref{GLE}] becomes
\begin{equation}
	\boldsymbol{\ddot r} =-\frac{1}{m}\int_0^t \gamma(t-t') \boldsymbol{v}(t')dt'-\omega^2 \boldsymbol{r} + \frac{1}{m}\boldsymbol{\eta}(t)+\frac{1}{m}\boldsymbol{\Omega}(t).
	\label{langavin_case2}
\end{equation}
To solve Eq.~\eqref{langavin_case2}, we define the response functions $\chi_1(t)$, $\chi_2(t)$, and $\chi_3(t)$ such that 
\begin{equation}
	\begin{aligned}
		\tilde\chi_1(s)=&\frac{\frac{\tilde\gamma(s)}{m}+s}{s^2+s\frac{\tilde\gamma(s)}{m}+\omega^2}, \quad \tilde\chi_2(s) = \frac{1}{s^2+s\frac{\tilde\gamma(s)}{m}+\omega^2},\\
		\text{and}\quad &
		\tilde\chi_3(s)=\frac{1}{1+\frac{1}{t_c}}.
	\end{aligned}
\end{equation}
Here, the tilde represents the Laplace transform.
With the help of these response functions, one can define the fluctuations in variables $\boldsymbol{r}$, $\boldsymbol{v}$, and $\boldsymbol{\Omega}$, which we denote by the parameters $\boldsymbol{g_1}(t)$, $\boldsymbol{g_2}(t)$, and $\boldsymbol{g_3}(t)$. The expressions for $\boldsymbol{g_i}(t)$'s are given by
\begin{align}
	\boldsymbol{g_1}(t)=&\boldsymbol{r}-\left[ \chi_1(t)\boldsymbol{r_0}+\chi_2(t)\boldsymbol{v_0}+\chi_{23}(t) \boldsymbol{\Omega_0}\right],
	\label{eq:g1_case2_def}\\
	\boldsymbol{g_2}(t)=&\boldsymbol{v}-\left[ \dot\chi_1(t)\boldsymbol{r_0}+\dot\chi_2(t)\boldsymbol{v_0}+\dot \chi_{23}(t) \boldsymbol{\Omega_0}\right],
	\label{eq:g2_case2_def}\\
	\text{and} \phantom{abc}& \nonumber\\
	\boldsymbol{g_3}(t)=&\boldsymbol{\Omega}-\chi_3(t) \boldsymbol{\Omega_0}.
	\label{eq:g3_case2_def}
\end{align}
After substituting expressions for $\boldsymbol{r}$, $\boldsymbol{v}$, and $\boldsymbol{\Omega}$ from Eqs.~\eqref{langavin_case2} and ~\eqref{active_dynamics} into Eqs.~\eqref{eq:g1_case2_def}-\eqref{eq:g3_case2_def}, we get
\begin{align}
	\boldsymbol{g_1}(t)=&\frac{1}{m}\int_0^t \chi_1(t')\boldsymbol{\eta}(t-t')dt' \nonumber
	\\
	&+\frac{D_\Omega}{m}\sqrt{\frac{2}{t_c} }\int_0^t \chi_{23}(t') \boldsymbol{\zeta}(t-t')dt',
\end{align}
\begin{align}
	\boldsymbol{g_2}(t)=&\frac{1}{m}\int_0^t \dot\chi_1(t')\boldsymbol{\eta}(t-t')dt' \nonumber
	\\
	&+\frac{D_\Omega}{m}\sqrt{\frac{2}{t_c} } \int_0^t \dot\chi_{23}(t') \boldsymbol{\zeta}(t-t')dt',
\end{align}
and
\begin{align}
	\boldsymbol{g_3}(t)=&D_\Omega \sqrt{\frac{2}{t_c}} \int_0^t \chi_3(t') \boldsymbol{\zeta}(t-t')dt',
\end{align}
with
\begin{equation}
	\chi_{23}(t)=(\chi_2 * \chi_3)(t)=\int_0^t \chi_2(t') e^{\frac{-(t-t')}{t_c}} dt'.
	\label{eq:chi23_case2}
\end{equation}
Taking the derivative of Eq.~\eqref{eq:chi23_case2} with respect to time, one can arrive at the following differential equation
\begin{equation}
	\dot\chi_{23}(t) = \chi_1(t) - \frac{1}{t_c} \chi_{23}(t).
\end{equation}
Now, we consider the second moments of fluctuations of $\boldsymbol{r}$, $\boldsymbol{v}$ and $\boldsymbol{\Omega}$. These moments can be written as a matrix $\Xi(t)$ with the elements $\Xi_{ij}=\langle g_i(t)
\cdot g_j(t)\rangle$. The diagonal components of the matrix $\Xi(t)$ are given by
\begin{equation}
	\Xi_{11}(t)=\frac{K_BT}{m^2} C_\eta^{(1)}(t) +\frac{2D_\Omega^2}{m^2t_c} \int_0^t \chi_{23}^2(t-t')dt',
\end{equation}
\begin{equation}
	\Xi_{22}(t)=\frac{K_BT}{m^2} C_\eta^{(2)}(t) +\frac{2D_\Omega^2}{m^2t_c} \int_0^t \dot\chi_{23}^2(t-t')dt',
\end{equation}
and
\begin{equation}
	\Xi_{33}(t)=\frac{2D_\Omega^2}{t_c} \int_0^t \chi_3^2(t-t')dt',
\end{equation}
where
\begin{equation}
	C_\eta^{(1)}(t) = \int_0^t dt'\int_0^tdt''\chi_{2}(t-t')\chi_{2}(t-t'')\gamma(|t'-t''|)
\end{equation}
and
\begin{equation}
	C_\eta^{(2)}(t) = \int_0^t dt'\int_0^tdt''\dot\chi_{2}(t-t')\dot\chi_{2}(t-t'')\gamma(|t'-t''|).
\end{equation}
Similarly, other components can be obtained as
\begin{equation}
	\Xi_{12}(t)=\Xi_{21}(t)=\frac{K_BT}{m^2} C_\eta^{(3)}(t) + \frac{D_\Omega^2}{m^2t_c} \left( \chi_{23}^2(t)-\chi_{23}^2(0) \right),
\end{equation}
\begin{equation}
	\Xi_{13}(t)=\Xi_{31}(t)=\frac{2D_\Omega^2}{t_c} \int_0^t \chi_{23}(t-t')\chi_3(t-t')dt',
\end{equation}
and
\begin{equation}
	\Xi_{23}(t)=\Xi_{32}(t)=\frac{2D_\Omega^2}{t_c} \int_0^t \dot\chi_{23}(t-t')\chi_3(t-t')dt'.
\end{equation}
Here, we have
\begin{equation}
	C_\eta^{(3)}(t) = \int_0^t dt'\int_0^tdt''\chi_{2}(t-t')\dot\chi_{2}(t-t'')\gamma(|t'-t''|).
\end{equation}
Due to the linearity of Eq.~\eqref{langavin_case2}, the probability distribution $P(\boldsymbol{\boldsymbol{r,v,\Omega}};t)$ is Gaussian and takes the form
\begin{equation}
	\begin{aligned}
		P(\boldsymbol{\boldsymbol{r,v,\Omega}};t)=\left(\frac{1}{2 \pi}\right)^\frac{3}{2} \frac{1}{\sqrt{|\Xi(t)|}}
		\exp\left[-\frac{1}{2} \boldsymbol{g}^T(t) \Xi^{-1}(t)\boldsymbol{g}(t)\right],
	\end{aligned}
	\label{eq:case2_pdf_non_markov}
\end{equation}
with
\begin{equation}
	\boldsymbol{g}(t)=\begin{bmatrix}
		\boldsymbol{g_1}(t)\\
		\boldsymbol{g_2}(t)\\
		\boldsymbol{g_3}(t)
	\end{bmatrix}.
\end{equation}
Hence, Eq.~\eqref{eq:case2_pdf_non_markov} represents the probability distribution function of an active Ornstein-Uhlenbeck particle confined by a harmonic confinement and suspended in a general non-Markovian media. In the next section, we discuss the non-Markovian dynamics of an active particle in the presence of an external magnetic field.

\subsection{ACTIVE PARTICLE UNDER EXTERNAL MAGNETIC FIELD}

Here, we consider the motion of an active Ornstein-Uhlenbeck particle suspended in a non-Markovian environment and subjected to the presence of an external magnetic field \(\mathbf{B} = (0,0,B_z)\). The magnetic Lorentz force \(\mathbf{F}_B = \frac{q}{m} (\mathbf{v} \times \mathbf{B})\) due to the magnetic field affects only in the \(x\)-\(y\) plane, leaving the \(z\)-direction unaffected. Thus, the motion remains confined to the \(x\)-\(y\) plane. The corresponding equation of motion [Eq.~\eqref{GLE}] is then given by 
\begin{align}
	m \dot{v_x} &= - \int_{0}^{t} \gamma(t - t') v_x(t') \, dt' + m\psi \dot{y} + \eta_x(t) + \Omega_x(t), \label{eq:model_case2_1}\\
	m \dot{v_y} &= - \int_{0}^{t} \gamma(t - t') v_y(t') \, dt' - m\psi \dot{x} + \eta_y(t) + \Omega_y(t),
	\label{eq:model_case2_2}
\end{align}
where \(\psi = \frac{q B}{m}\) is the cyclotron frequency. First, we consider the non-Markovian medium as a special case of Jeffrey fluid. Hence, using the Jeffrey fluid model [Eq.~\eqref{Jeff_fluid}],
one can write the dynamics Eqs.~\eqref{eq:model_case2_1} and \eqref{eq:model_case2_2} in matrix form as in Eq.~\eqref{eq:dyn_matrix_form}, with the matrices $\bm{X}$, $A$, and $B$ given by
\begin{widetext}
	\begin{equation}
		\mathbf{X} =
		\begin{bmatrix}
			v_x & v_y & \xi_x & \xi_y & \eta_{2x}&\eta_{2y} & \Omega_{x}&\Omega_{y}    \end{bmatrix}^{T}.
	\end{equation}
	\begin{equation}
		A =
		\begin{bmatrix}
			-\frac{\gamma_f}{2m} & \psi & -\frac{\gamma_s}{2m} & 0 & \frac{1}{m} & 0 & \frac{1}{m} & 0 \\
			-\psi & -\frac{\gamma_f}{2m} & 0 & -\frac{\gamma_s}{2m} & 0 & \frac{1}{m} & 0 & \frac{1}{m} \\
			\frac{1}{t_{s}} & 0 & -\frac{1}{t_{s}} & 0 & 0 & 0 & 0 & 0 \\
			0 & \frac{1}{t_{s}} & 0 & -\frac{1}{t_{s}} & 0 & 0 & 0 & 0 \\
			0 & 0 & 0 & 0 & -\frac{1}{t_{s}} & 0 & 0 & 0 \\
			0 & 0 & 0 & 0 & 0 & -\frac{1}{t_{s}} & 0 & 0 \\
			0 & 0 & 0 & 0 & 0 & 0 & -\frac{1}{t_{c}} & 0 \\
			0 & 0 & 0 & 0 & 0 & 0 & 0 & -\frac{1}{t_{c}}
		\end{bmatrix}
		,
		\quad
		B =
		\begin{bmatrix}
			0 & 0 & \frac{1}{m^2} & 0 & 0 & 0 & 0 & 0 & 0 & 0 \\
			0 & 0 & 0 & \frac{1}{m^2} & 0 & 0 & 0 & 0 & 0 & 0 \\
			0 & 0 & 0 & 0 & 0 & 0 & 0 & 0 & 0 & 0 \\
			0 & 0 & 0 & 0 & 0 & 0 & 0 & 0 & 0 & 0 \\
			0 & 0 & 0 & 0 & 0 & 0 & \frac{k_B T \gamma_s}{t_{s}} & 0 & 0 & 0 \\
			0 & 0 & 0 & 0 & 0 & 0 & 0 & \frac{k_B T \gamma_s}{t_{s}} & 0 & 0 \\
			0 & 0 & 0 & 0 & 0 & 0 & 0 & 0 & \frac{2 \xi_0^2}{t_{c}} & 0 \\
			0 & 0 & 0 & 0 & 0 & 0 & 0 & 0 & 0 & \frac{2 \xi_0^2}{t_{c}}
		\end{bmatrix}
		.
	\end{equation}
	Now, the corresponding Fokker-Planck equation takes the form
		\begin{equation}
			\begin{aligned}
				\frac{\partial P}{\partial t} &=  \frac{\gamma_f}{2m} \frac{\partial}{\partial v_x} (v_x P)
				+ \frac{\gamma_f}{2m} \frac{\partial}{\partial v_y} (v_y P)  +\left(\frac{\gamma_s}{2m} \xi_x - \psi v_y - \frac{1}{m} \eta_{2x} - \frac{1}{m} \Omega_{x} \right) \frac{\partial P}{\partial v_x} \\
				&\quad + \left(\frac{\gamma_s}{2m} \xi_y + \psi v_x - \frac{1}{m} \eta_{2y} - \frac{1}{m} \Omega_{y} \right) \frac{\partial P}{\partial v_y} - \frac{v_x}{t_{s}} \frac{\partial P}{\partial \xi_x} 
				+ \frac{1}{t_{s}} \frac{\partial}{\partial \xi_x} (\xi_x P) 
				- \frac{v_y}{t_{s}} \frac{\partial P}{\partial \xi_y} \\
				&\quad
				+ \frac{1}{t_{s}} \frac{\partial}{\partial \xi_y} (\xi_y P)  + \frac{1}{t_{s}} \frac{\partial}{\partial \eta_{2x}} (\eta_{2x} P)
				+ \frac{1}{t_{s}} \frac{\partial}{\partial \eta_{2y}} (\eta_{2y} P) 
				+ \frac{1}{t_{c}} \left( \frac{\partial}{\partial \Omega_{x}} (\Omega_{x} P) 
				+  \frac{\partial}{\partial \Omega_{y}} (\Omega_{y} P) \right) \\
				&\quad + \frac{1}{2m^2} \left( \frac{\partial^2 P}{\partial v_x^2} 
				+  \frac{\partial^2 P}{\partial v_y^2} \right)
				+ \frac{k_B T \gamma_s}{2 t_{s}} \left( \frac{\partial^2 P}{\partial \eta_{2x}^2} 
				+  \frac{\partial^2 P}{\partial \eta_{2y}^2} \right) + \frac{2 \xi_0^2}{t_{c}} \left( \frac{\partial^2 P}{\partial \Omega_{x}^2}
				+\frac{\partial^2 P}{\partial \Omega_{y}^2} \right) .
			\end{aligned}
		\end{equation}    
\end{widetext}

Next, we consider the case of a general non-Markovian kernel $\gamma(t-t')$ which captures the feature of a general non-Markovian medium. Then, the equation of motion [Eq.~\eqref{GLE}] along with Eq.~\eqref{active_dynamics} is given by
\begin{align}
	\dot v_x(t)&=-\frac{1}{m}\int_0^t\gamma(t-t')v_x(t')dt'+ \psi v_y(t) \nonumber
	\\
	+&\frac{1}{m}\eta_x(t)+\frac{1}{m}\Omega_x(t) \label{langavin_case3_1}\\
	\dot v_y(t)&=-\frac{1}{m}\int_0^t\gamma(t-t')v_y(t')dt'- \psi v_x(t) \nonumber
	\\
	+&\frac{1}{m}\eta_y(t)+\frac{1}{m}\Omega_y(t) \label{langavin_case3_2}\\
	\dot\Omega_x(t)&=-\frac{1}{t_c}\Omega_x(t)+\frac{D_\Omega}{t_c} \zeta_x(t) \label{langavin_case3_3}\\
	\dot\Omega_y(t)&=-\frac{1}{t_c}\Omega_y(t)+\frac{D_\Omega}{t_c} \zeta_y(t).
	\label{langavin_case3_4}
\end{align}
Following the same procedure as before, we define the fluctuation terms as
\begin{equation}
	\begin{aligned}
		g_1(t)=&v_x-[\chi_1(t)v_{x0}+\chi_2(t)v_{y0}+\chi_{23}(t)\Omega_{y0}\\
		+&\chi_{13}(t)\Omega_{x0}],
	\end{aligned}
\end{equation}
\begin{equation}
	\begin{aligned}
		g_2(t)=&v_y-[ \chi_1(t)v_{y0}-\chi_2(t)v_{x0}-\chi_{23}(t)\Omega_{x0}\\
		+&\chi_{13}(t)\Omega_{y0} ],
	\end{aligned}
\end{equation}
\begin{equation}
	\begin{aligned}
		g_3(t)=&\Omega_x-\chi_3(t) \Omega_{x0},
	\end{aligned}
\end{equation}
and
\begin{equation}
	\begin{aligned}
		g_4(t)=&\Omega_y-\chi_3(t) \Omega_{y0}.
	\end{aligned}
\end{equation}
Here, the response functions $\chi_1(t)$, $\chi_2(t)$, and $\chi_3(t)$ are defined by their Laplace transforms
\begin{equation}
	\begin{aligned}
		\tilde\chi_1(s)&=\frac{\left(s+\frac{\tilde\gamma(s)}{m}\right)}{\left(s+\frac{\tilde\gamma(s)}{m}\right)^2 + \psi^2 },
	\end{aligned}  
\end{equation}
\begin{equation}
	\begin{aligned}
		\tilde\chi_2(s)=\frac{\psi}{\left(s+\frac{\tilde\gamma(s)}{m}\right)^2 + \psi^2 },
	\end{aligned}  
\end{equation}
and
\begin{equation}
	\begin{aligned}
		\tilde \chi_3(s) &= \frac{1}{s+\frac{1}{t_c}}.
	\end{aligned}  
\end{equation}
Also,
\begin{equation}
	\chi_{13}(t)=(\chi_1 * \chi_3)(t)=\int_0^t \chi_1(t') e^{\frac{-(t-t')}{t_c}} dt',
\end{equation}
and 
\begin{equation}
	\chi_{23}(t)=(\chi_2 * \chi_3)(t)=\int_0^t \chi_2(t') e^{\frac{-(t-t')}{t_c}} dt'.
\end{equation}
Thus, one can express $g_1(t)$, $g_2(t)$, $g_3(t)$, and $g_4(t)$ as
\begin{widetext}
	\begin{equation}
		\begin{aligned}
			g_1(t)=&\int_0^t \chi_2(t')\eta_y(t-t')dt'+\frac{D_\Omega}{t_c}\int_0^t \chi_{23}(t')\zeta_y(t-t')dt'+
			\int_0^t \chi_1(t')\eta_x(t-t')dt'+\frac{D_\Omega}{t_c}\int_0^t \chi_{12}(t')\zeta_x(t-t')dt',\\
			g_2(t)=&\int_0^t \chi_2(t')\eta_x(t-t')dt'-\frac{D_\Omega}{t_c}\int_0^t \chi_{23}(t')\zeta_x(t-t')dt'+
			\int_0^t \chi_1(t')\eta_y(t-t')dt'+\frac{D_\Omega}{t_c}\int_0^t \chi_{12}(t')\zeta_y(t-t')dt',\\
			g_3(t)=&\frac{D_\Omega}{t_c} \int_0^t \chi_3(t) \zeta_x(t-t')dt',\\
			g_4(t)=&\frac{D_\Omega}{t_c} \int_0^t \chi_3(t) \zeta_y(t-t')dt'.
		\end{aligned}
	\end{equation}
	
\end{widetext}

Now, we consider all the second moments represented by the matrix $\Xi(t)$ with matrix elements $\Xi_{ij}=\langle g_i(t) \cdot g_j(t)\rangle$. The elements of these matrices are given by
\begin{equation}
	\begin{aligned}
		\Xi_{11}(t)&=\Xi_{22}(t) = \frac{K_BT}{m^2} C_\eta^{(1)}(t)+\frac{K_BT}{m^2} C_\eta^{(2)}(t)\\
		&+\frac{2D_\Omega^2}{t_c} \int_0^t \chi_{23}^2(t-t')dt'
		+\frac{2D_\Omega^2}{t_c} \int_0^t \chi_{13}^2(t-t')dt',
	\end{aligned}
\end{equation}
\begin{equation}
	\begin{aligned}
		\Xi_{33}(t)=\Xi_{44}(t) = &\frac{2D_\Omega^2}{t_c} \int_0^t \chi_{3}^2(t-t')dt',
	\end{aligned}
\end{equation}
\begin{equation}
	\begin{aligned}
		\Xi_{13}(t) &= \Xi_{31}(t)= \Xi_{24}(t)=\Xi_{42}(t)\\
		&=\frac{2D_\Omega^2}{m t_c}\int_0^t \chi_{13}(t-t')\chi_{3}(t-t')dt',
	\end{aligned}
\end{equation}
\begin{equation}
	\begin{aligned}
		\Xi_{14}(t) &= \Xi_{41}(t)= \Xi_{23}(t)=\Xi_{32}(t) \\
		&= \frac{2D_\Omega^2}{m t_c}\int_0^t \chi_{23}(t-t')\chi_{3}(t-t')dt',
	\end{aligned}
\end{equation}
and
\begin{equation}
	\begin{aligned}
		&\Xi_{12}(t)=\Xi_{21}(t)=\Xi_{34}(t)=\Xi_{43}(t)=0,
	\end{aligned}
\end{equation}
where
\begin{equation}
	C_\eta^{(1)}(t) = \int_0^t dt'\int_0^tdt''\chi_{2}(t-t')\chi_{2}(t-t'')\gamma(|t'-t''|)
\end{equation}
and
\begin{equation}
	C_\eta^{(2)}(t) = \int_0^t dt'\int_0^tdt''\chi_{1}(t-t')\chi_{1}(t-t'')\gamma(|t'-t''|).
\end{equation}
Thus, the probability distribution function associated with Eqs.~\eqref{langavin_case3_1}-\eqref{langavin_case3_4} is given by
\begin{equation}
	\begin{aligned}
		P(v_x,v_y,\Omega_x,\Omega_y;t) &=
		\left(\frac{1}{2 \pi}\right)^2 \frac{1}{\sqrt{|\Xi(t)|}} \\
		&\times
		\exp\left[-\frac{1}{2} \boldsymbol{g}^T(t) \Xi^{-1}(t)\boldsymbol{g}(t)\right],
	\end{aligned}
	\label{eq:case3_pdf_non_markov}
\end{equation}
with
\begin{equation}
	\boldsymbol{g}(t)=\begin{bmatrix}
		\boldsymbol{g_1}(t)\\
		\boldsymbol{g_2}(t)\\
		\boldsymbol{g_3}(t)\\
		\boldsymbol{g_4}(t)
	\end{bmatrix}.
\end{equation}
Hence, Eq.~\eqref{eq:case3_pdf_non_markov} represents the probability distribution function of an active Ornstein-Uhlenbeck particle in the presence of an external magnetic field and suspended in a general non-Markovian environment. It would be further interesting to examine the combined effect of harmonic confinement and a magnetic field. In the next section, we discuss this aspect of the dynamics.

\subsection{ACTIVE PARTICLE WITH HARMONIC CONFINEMENT UNDER EXTERNAL MAGNETIC FIELD}
Here, we consider a confined harmonic particle in the presence of an external magnetic field and suspended in a non-Markovian environment. The corresponding equation of motion [Eq.~\eqref{GLE}] becomes
\begin{equation}
	m \ddot{x} = - \int_{0}^{t} \gamma(t - t') \dot{x}(t') \, dt' + m\psi \dot{y} - m\omega^2 x + \eta_x(t)+ \Omega_x(t),
	\label{langevin_case4_x}
\end{equation}
and
\begin{equation}
	m \ddot{y} = - \int_{0}^{t} \gamma(t - t') \dot{y}(t') \, dt' - m\psi \dot{x} - m\omega^2 y + \eta_y(t)+\Omega_y(t),
	\label{langevin_case4_y}
\end{equation}
where the cyclotron frequency $\psi = \frac{qB}{m}$ and $\omega$ is the harmonic frequency. As discussed in the previous cases, we first consider the non-Markovian environment as a special case of Jefferey fluid. Hence, using Jeffreys fluid framework, for describing the motion of the particle, the corresponding expressions of the matrices $\bm{X}$, $A$, and $B$ in Eq.~\eqref{eq:dyn_matrix_form} takes the form
	\begin{equation}
		\mathbf{X} =
		\begin{bmatrix}
			x & y & v_x & v_y & \xi_x & \xi_y & \eta_{2x}&\eta_{2y} & \Omega_{x}&\Omega_{y}    \end{bmatrix}^{T},
	\end{equation}
	\begin{equation}
		A =
		\begin{bmatrix}
			0 & 0 & 1 & 0 & 0 & 0 & 0 & 0 & 0 & 0 \\
			0 & 0 & 0 & 1 & 0 & 0 & 0 & 0 & 0 & 0 \\
			-\omega^2 & 0 & -\frac{\gamma_f}{2m} & \psi & -\frac{\gamma_s}{2m} & 0 & \frac{1}{m} & 0 & \frac{1}{m} & 0 \\
			0 & -\omega_0^2 & -\psi & -\frac{\gamma_f}{2m} & 0 & -\frac{\gamma_s}{2m} & 0 & \frac{1}{m} & 0 & \frac{1}{m} \\
			0 & 0 & \frac{1}{t_s} & 0 & -\frac{1}{t_s} & 0 & 0 & 0 & 0 & 0 \\
			0 & 0 & 0 & \frac{1}{t_s} & 0 & -\frac{1}{t_s} & 0 & 0 & 0 & 0 \\
			0 & 0 & 0 & 0 & 0 & 0 & -\frac{1}{t_{s}} & 0 & 0 & 0 \\
			0 & 0 & 0 & 0 & 0 & 0 & 0 & -\frac{1}{t_{s}} & 0 & 0 \\
			0 & 0 & 0 & 0 & 0 & 0 & 0 & 0 & -\frac{1}{t_{c}} & 0 \\
			0 & 0 & 0 & 0 & 0 & 0 & 0 & 0 & 0 & -\frac{1}{t_{c}}
		\end{bmatrix}
		,
	\end{equation}
	and
	\begin{equation}
		B =
		\begin{bmatrix}
			0 & 0 & 0 & 0 & 0 & 0 & 0 & 0 & 0 & 0 \\
			0 & 0 & 0 & 0 & 0 & 0 & 0 & 0 & 0 & 0 \\
			0 & 0 & \frac{1}{m^2} & 0 & 0 & 0 & 0 & 0 & 0 & 0 \\
			0 & 0 & 0 & \frac{1}{m^2} & 0 & 0 & 0 & 0 & 0 & 0 \\
			0 & 0 & 0 & 0 & 0 & 0 & 0 & 0 & 0 & 0 \\
			0 & 0 & 0 & 0 & 0 & 0 & 0 & 0 & 0 & 0 \\
			0 & 0 & 0 & 0 & 0 & 0 & \frac{k_B T \gamma_s}{t_{s1}} & 0 & 0 & 0 \\
			0 & 0 & 0 & 0 & 0 & 0 & 0 & \frac{k_B T \gamma_s}{t_{s1}} & 0 & 0 \\
			0 & 0 & 0 & 0 & 0 & 0 & 0 & 0 & \frac{2 \xi_0^2}{t_{s2}} & 0 \\
			0 & 0 & 0 & 0 & 0 & 0 & 0 & 0 & 0 & \frac{2 \xi_0^2}{t_{s2}}
		\end{bmatrix}.
	\end{equation}
	Now, the corresponding Fokker-Planck equation is given by
	\begin{widetext}
		\begin{equation}
			\begin{aligned}
				\frac{\partial P}{\partial t} &= 
				- v_x \frac{\partial P}{\partial x} 
				- v_y \frac{\partial P}{\partial y} 
				+ \frac{\gamma_f}{2m} \frac{\partial}{\partial v_x} (v_x P)
				+ \frac{\gamma_f}{2m} \frac{\partial}{\partial v_y} (v_y P)+\left(  - \psi v_y + \frac{\gamma_s}{2m} \xi_x - \frac{1}{m} \eta_{2x} - \frac{1}{m} \Omega_{x} - \omega_0^2 x \right) \frac{\partial P}{\partial v_x} \\
				&\quad + \left( \psi v_x + \frac{\gamma_s}{2m} \xi_y - \frac{1}{m} \eta_{2y} - \frac{1}{m} \Omega_{y} - \omega_0^2 y \right) \frac{\partial P}{\partial v_y}  - \frac{v_x}{t_s} \frac{\partial P}{\partial \xi_x} 
				+ \frac{1}{t_s} \frac{\partial}{\partial \xi_x} (\xi_x P) 
				- \frac{v_y}{t_s} \frac{\partial P}{\partial \xi_y} 
				+ \frac{1}{t_s} \frac{\partial}{\partial \xi_y} (\xi_y P) \\
				&\quad + \frac{1}{t_{s}} \frac{\partial}{\partial \eta_{2x}} (\eta_{2x} P) 
				+ \frac{1}{t_{s1}} \frac{\partial}{\partial \eta_{2y}} (\eta_{2y} P) 
				+ \frac{1}{t_{c}} \left( \frac{\partial}{\partial \Omega_{x}} (\Omega_{x} P)
				+ \frac{\partial}{\partial \Omega_{y}} (\Omega_{y} P) \right)  + \frac{1}{2m^2} \left( \frac{\partial^2 P}{\partial v_x^2} 
				+  \frac{\partial^2 P}{\partial v_y^2} \right) \\
				&\quad 
				+ \frac{k_B T \gamma_s}{2 t_{s}} \left( \frac{\partial^2 P}{\partial \eta_{2x}^2} 
				+ \frac{\partial^2 P}{\partial \eta_{2y}^2} \right) + \frac{2 \xi_0^2}{t_{c}} \left( \frac{\partial^2 P}{\partial \Omega_{x}^2}
				+ \frac{\partial^2 P}{\partial \Omega_{y}^2} \right) .
			\end{aligned}
		\end{equation}    
	\end{widetext}
	
	Next, we consider the case of a general non-Markovian medium, characterized by the friction kernel $\gamma(t-t')$. Following the previously outlined cases, the fluctuation terms can be expressed as
	\begin{equation}
		\begin{aligned}
			g_1(t)=x-[ \chi_1(t)x_0+\chi_2(t)y_0+\chi_4(t)v_{x0}+\chi_3(t)v_{y0}
			\\
			+\chi_{45}(t) \Omega_{x0}+\chi_{35}(t) \Omega_{y0} ],
		\end{aligned}
		\label{g1_def_case4}
	\end{equation}
	\begin{equation}
		\begin{aligned}
			g_2(t)=y-[ \chi_1(t)y_0-\chi_2(t)x_0+\chi_4(t)v_{y0}-\chi_3(t)v_{x0}
			\\
			+\chi_{45}(t) \Omega_{y0}-\chi_{35}(t) \Omega_{x0} ],
		\end{aligned}
	\end{equation}
	\begin{equation}
		\begin{aligned}
			g_3(t)=v_x-[ \dot\chi_1(t)x_0+\dot\chi_2(t)y_0+\dot\chi_4(t)v_{x0}+\dot\chi_3(t)v_{y0}
			\\
			+\dot\chi_{45}(t) \Omega_{x0}+\dot\chi_{35}(t) \Omega_{y0}],
		\end{aligned}
	\end{equation}
	\begin{equation}
		\begin{aligned}
			g_4(t)=v_y-[ \dot\chi_1(t)y_0-\dot\chi_2(t)x_0+\dot\chi_4(t)v_{y0}-\dot\chi_3(t)v_{x0}
			\\
			+\dot\chi_{45}(t) \Omega_{y0}-\dot\chi_{35}(t) \Omega_{x0}],
		\end{aligned}
	\end{equation}
	\begin{equation}
		\begin{aligned}
			g_5(t)=&\Omega_x-\chi_5(t) \Omega_{0x},
		\end{aligned}
	\end{equation}
	and
	\begin{equation}
		\begin{aligned}
			g_6(t)=&\Omega_y-\chi_5(t) \Omega_{0y}.
		\end{aligned}
		\label{g6_def_case4}
	\end{equation}
	The corresponding response functions are given by their Laplace transform
	\begin{equation}
		\tilde\chi_1(s)=\frac{\left( s^2+\frac{\tilde\gamma(s)}{m} s+\omega^2 \right) \left( \frac{\tilde\gamma(s)}{m} +s \right)+s\psi^2}{\left( s^2+\frac{\tilde\gamma(s)}{m} s+\omega^2 \right)^2 + s^2 \psi^2},
	\end{equation}
	\begin{equation}
		\tilde\chi_2(s)=\frac{s\psi\left( \frac{\tilde\gamma(s)}{m} + s\right) - \psi \left( s^2+\frac{\tilde\gamma(s)}{m} s+\omega^2 \right)}{\left( s^2+\frac{\tilde\gamma(s)}{m} s+\omega^2 \right)^2 + s^2 \psi^2},
	\end{equation}
	\begin{equation}
		\tilde\chi_3(s)=\frac{s\psi}{\left( s^2+\frac{\tilde\gamma(s)}{m} s+\omega^2 \right)^2 + s^2 \psi^2},
	\end{equation}
	\begin{equation}
		\tilde\chi_4(s)=\frac{s^2+\frac{\tilde\gamma(s)}{m} s+\omega^2}{\left( s^2+\frac{\tilde\gamma(s)}{m} s+\omega^2 \right)^2 + s^2 \psi^2},
	\end{equation}
	and
	\begin{equation}
		\tilde\chi_5(s)=\frac{1}{s+\frac{1}{t_c}}.
	\end{equation}
	\begin{widetext}
		Also, we consider $\chi_{ij}(t) = (\chi_i * \chi_j)(t)$.
		Substituting the solution of Eqs.~\eqref{langevin_case4_x} and \eqref{langevin_case4_y} into Eqs.~\eqref{g1_def_case4}-\eqref{g6_def_case4}, we get the expressions for $g_i$'s as
		\begin{equation}
			\begin{aligned}
				g_1(t)=\frac{1}{m} \int_0^t \chi_4(t-t')\eta_x(t')dt'+\frac{1}{m} \int_0^t \chi_3(t-t')\eta_y(t')dt'+\frac{D_\Omega}{m}\sqrt{\frac{2}{t_c}} \int_0^t \chi_{45}(t-t')\zeta_x(t')dt'
				\\
				+\frac{D_\Omega}{m}\sqrt{\frac{2}{t_c}} \int_0^t \chi_{35}(t-t')\zeta_y(t')dt',
			\end{aligned}
		\end{equation}
		\begin{equation}
			\begin{aligned}
				g_2(t)=\frac{1}{m} \int_0^t \chi_4(t-t')\eta_y(t')dt'+\frac{1}{m} \int_0^t \chi_3(t-t')\eta_x(t')dt'+\frac{D_\Omega}{m}\sqrt{\frac{2}{t_c}} \int_0^t \chi_{45}(t-t')\zeta_y(t')dt'
				\\
				+\frac{D_\Omega}{m}\sqrt{\frac{2}{t_c}} \int_0^t \chi_{35}(t-t')\zeta_x(t')dt',
			\end{aligned}
		\end{equation}
		\begin{equation}
			\begin{aligned}
				g_3(t)=\frac{1}{m} \int_0^t \dot\chi_4(t-t')\eta_x(t')dt'+\frac{1}{m} \int_0^t \dot\chi_3(t-t')\eta_y(t')dt'+\frac{D_\Omega}{m}\sqrt{\frac{2}{t_c}} \int_0^t \dot\chi_{45}(t-t')\zeta_x(t')dt'
				\\
				+\frac{D_\Omega}{m}\sqrt{\frac{2}{t_c}} \int_0^t \dot\chi_{35}(t-t')\zeta_y(t')dt',
			\end{aligned}
		\end{equation}
		\begin{equation}
			\begin{aligned}
				g_4(t)=\frac{1}{m} \int_0^t \dot\chi_4(t-t')\eta_y(t')dt'+\frac{1}{m} \int_0^t \dot\chi_3(t-t')\eta_x(t')dt'+\frac{D_\Omega}{m}\sqrt{\frac{2}{t_c}} \int_0^t \dot\chi_{45}(t-t')\zeta_y(t')dt'
				\\
				+\frac{D_\Omega}{m}\sqrt{\frac{2}{t_c}} \int_0^t \dot\chi_{35}(t-t')\zeta_x(t')dt',
			\end{aligned}
		\end{equation}
		\begin{equation}
			\begin{aligned}
				g_5(t)=&D_\Omega \sqrt{\frac{2}{t_c}} \int_0^t \chi_5(t-t') \zeta_x(t')dt',
			\end{aligned}
		\end{equation}
		and
		\begin{equation}
			\begin{aligned}
				g_5(t)=&D_\Omega \sqrt{\frac{2}{t_c}} \int_0^t \chi_5(t-t') \zeta_y(t')dt'.
			\end{aligned}
		\end{equation}
	\end{widetext}
	Now, we consider all the second moments represented by the matrix $\Xi(t)$ with $\Xi_{ij}=\langle g_i(t) \cdot g_j(t)\rangle$. The elements of this matrix are given by
	\begin{equation}
		\begin{aligned}
			\Xi_{11}(t)&=\Xi_{22}(t) = \frac{K_BT}{m^2} C_\eta^{(1)}(t)+\frac{K_BT}{m^2} C_\eta^{(2)}(t)\\
			&+\frac{2D_\Omega^2}{t_c} \int_0^t \chi_{45}^2(t-t')dt'
			+\frac{2D_\Omega^2}{t_c} \int_0^t \chi_{35}^2(t-t')dt',
		\end{aligned}
	\end{equation}
	\begin{equation}
		\begin{aligned}
			\Xi_{33}(t)&=\Xi_{44}(t) = \frac{K_BT}{m^2} C_\eta^{(3)}(t)+\frac{K_BT}{m^2} C_\eta^{(4)}(t)\\
			&+\frac{2D_\Omega^2}{m^2t_c} \int_0^t \dot\chi_{45}^2(t-t')dt'
			+\frac{2D_\Omega^2}{m^2t_c} \int_0^t \dot\chi_{35}^2(t-t')dt',
		\end{aligned}
	\end{equation}
	\begin{equation}
		\Xi_{55}(t)=\Xi_{66}(t) = \frac{2D_\Omega^2}{t_c} \int_0^t \chi_5^2(t-t')dt',
	\end{equation}
	\begin{equation}
		\begin{aligned}
			\Xi_{12}(t)=\Xi_{21}(t)=\Xi_{34}(t)=\Xi_{43}(t)=\Xi_{56}(t)=\Xi_{65}(t)=0,
		\end{aligned}
	\end{equation}
	\begin{equation}
		\begin{aligned}
			\Xi_{13}(t)&=\Xi_{31}(t)=\Xi_{24}(t)=\Xi_{42}(t)\\
			&= \frac{K_BT}{m^2} C_\eta^{(5)}(t)+\frac{K_BT}{m^2} C_\eta^{(6)}(t)\\
			&+\frac{D_\Omega^2}{t_c} (\chi_{45}^2(t)-\chi_{45}^2(0))
			+\frac{2D_\Omega^2}{t_c} (\chi_{35}^2(t)-\chi_{35}^2(0)),
		\end{aligned}
	\end{equation}
	\begin{equation}
		\begin{aligned}
			\Xi_{14}(t)&=\Xi_{41}(t)=-\Xi_{23}(t)=-\Xi_{32}(t)\\
			&= \frac{K_BT}{m^2} C_\eta^{(7)}(t)+\frac{K_BT}{m^2} C_\eta^{(8)}(t)\\
			&+\frac{D_\Omega^2}{t_c} \int_0^t W_{\chi_{35},\chi_{45}}(t-t')dt',
		\end{aligned}
	\end{equation}
	\begin{equation} 
		\begin{aligned}
			\Xi_{15}(t)&=\Xi_{51}(t)=\Xi_{26}(t)=\Xi_{62}(t)\\
			&=\frac{2D_\Omega^2}{mt_c} \int_0^t \chi_{45}(t-t')\chi_5(t-t')dt',
		\end{aligned}
	\end{equation}
	\begin{equation}
		\begin{aligned}
			\Xi_{16}(t) &= \Xi_{61}(t)=-\Xi_{25}(t)=-\Xi_{52}(t)\\
			&= \frac{2D_\Omega^2}{mt_c} \int_0^t \chi_{35}(t-t')\chi_5(t-t')dt',
		\end{aligned}
	\end{equation}
	\begin{equation}
		\begin{aligned}
			\Xi_{35}(t) &= \Xi_{53}(t)=\Xi_{46}(t)=\Xi_{64}(t)
			\\
			&=\frac{2D_\Omega^2}{mt_c} \int_0^t \dot\chi_{45}(t-t')\chi_5(t-t')dt',
		\end{aligned}
	\end{equation}
	and
	\begin{equation}
		\begin{aligned}
			\Xi_{36}(t) &= \Xi_{63}(t)=-\Xi_{45}(t)=-\Xi_{54}(t)=
			\\
			&= \frac{2D_\Omega^2}{mt_c} \int_0^t \dot\chi_{35}(t-t')\chi_5(t-t')dt',
		\end{aligned}
	\end{equation}
	where
	\begin{equation}
		C_\eta^{(1)}(t) = \int_0^t dt'\int_0^tdt''\chi_{4}(t-t')\chi_{4}(t-t'')\gamma(|t'-t''|),
	\end{equation}
	\begin{equation}
		C_\eta^{(2)}(t) = \int_0^t dt'\int_0^tdt''\chi_{3}(t-t')\chi_{3}(t-t'')\gamma(|t'-t''|),
	\end{equation}
	\begin{equation}
		C_\eta^{(3)}(t) = \int_0^t dt'\int_0^tdt''\dot\chi_{4}(t-t')\dot\chi_{4}(t-t'')\gamma(|t'-t''|),
	\end{equation}
	\begin{equation}
		C_\eta^{(4)}(t) = \int_0^t dt'\int_0^tdt''\dot\chi_{3}(t-t')\dot\chi_{3}(t-t'')\gamma(|t'-t''|),
	\end{equation}
	\begin{equation}
		C_\eta^{(5)}(t) = \int_0^t dt'\int_0^tdt''\chi_{4}(t-t')\dot\chi_{4}(t-t'')\gamma(|t'-t''|),
	\end{equation}
	\begin{equation}
		C_\eta^{(6)}(t) = \int_0^t dt'\int_0^tdt''\chi_{3}(t-t')\dot\chi_{3}(t-t'')\gamma(|t'-t''|),
	\end{equation}
	and
	\begin{equation}
		C_\eta^{(7)}(t) = \int_0^t dt'\int_0^tdt''\chi_{4}(t-t')\dot\chi_{3}(t-t'')\gamma(|t'-t''|).
	\end{equation}
	Also, the Wronskian
	\begin{equation}
		\quad W_{\chi_{35},\chi_{45}}=\chi_{35}(t-t')\dot\chi_{45}(t-t')-\chi_{45}(t-t')\dot\chi_{35}(t-t').
	\end{equation}
	Now, due to the linearity of Eqs.~\eqref{langevin_case4_x} and ~\eqref{langevin_case4_y}, the corresponding probability distribution function takes a Gaussian form and is given by
	\begin{equation}
		\begin{aligned}
			P(x,y,v_x,v_y,\Omega_x,\Omega_y;t) &=
			\left(\frac{1}{2 \pi}\right)^3 \frac{1}{\sqrt{|\Xi(t)|}}\\
			&\times
			\exp\left[-\frac{1}{2} \boldsymbol{g}^T(t) \Xi^{-1}(t)\boldsymbol{g}(t)\right]
		\end{aligned}
		\label{eq:case4_pdf_non_markov}
	\end{equation}
	with
	\begin{equation}
		\boldsymbol{g}(t)=\begin{bmatrix}
			\boldsymbol{g_1}(t)\\
			\boldsymbol{g_2}(t)\\
			\boldsymbol{g_3}(t)\\
			\boldsymbol{g_4}(t)\\
			\boldsymbol{g_5}(t)\\
			\boldsymbol{g_6}(t)
		\end{bmatrix}.
	\end{equation}
	
	Hence, Eq.~\eqref{eq:case4_pdf_non_markov} represents the probability distribution function of an harmonically confined active Ornstein-Uhlenbeck particle in the presence of an external magnetic field and suspended in a general non-Markovian environment.

\section{CONCLUSIONS}\label{sec:summary}

In this work, we have analyzed the non-Markovian dynamics of an active Ornstein-Uhlenbeck particle by formulating the corresponding Fokker-Planck equations in various physical settings. Four distinct cases were considered: (i) the free active particle, 
(ii) an active particle subjected to harmonic confinement, 
 (iii) an active particle under the influence of a magnetic field, and (iv) an active particle simultaneously experiencing both harmonic confinement and a magnetic field.
For each configuration, the Fokker-Planck equation was first established explicitly for the Jeffrey fluid model, which represents a special case of non-Markovian memory. Building on this, we extended the analysis to general memory kernels and derived the associated probability distributions. In the case of the free active particle, we further obtained the complete Fokker-Planck equation valid for an arbitrary memory kernel, thereby providing the most general description. Importantly, our results reduce smoothly to familiar Brownian Markovian expressions in the appropriate limits~\cite{adelman1976fokker}, confirming the consistency of the approach.
To the best of our knowledge, this is the first attempt to develop such a general formalism for active particles in the presence of non-Markovian memory effects. 
The methodology presented here may find applications in a variety of contexts where active particles operate in complex environments. Examples include intracellular transport in viscoelastic cytoplasm, active colloids suspended in structured fluids, and microswimmers influenced by electromagnetic fields or mechanical confinement.

The framework established in this study highlights the importance of incorporating non-Markovian memory in the theoretical treatment of active particles. By demonstrating both mathematical consistency and physical applicability, our results provide a foundation for future studies aiming to bridge microscopic dynamics with macroscopic observables in active matter systems under realistic non-Markovian conditions.

\section*{ACKNOWLEDGEMENT}
MS acknowledges the computational facility, Department of Physics, University of Kerala, SERB-SURE grant (SUR/2022/000377), and CRG grant (CRG/2023/002026) by the Department of Science and Technology, Government of India, for financial support.
\vspace{2em}

SSP and MM contributed equally to this work.

\section*{DATA AVAILABILITY}
The data that support the findings of this study are available within the article.

\end{document}